\documentclass[twocolumn, trackchanges]{aastex62}

\graphicspath{{./}{figures/}}

\shorttitle{The Landscape of CL AGN Host Galaxies}
\shortauthors{Dodd et al.}

\begin{document}

\title{The Landscape of Galaxies Harboring Changing-Look Active Galactic Nuclei in the Local Universe}

\author[0000-0002-3696-8035]{Sierra A. Dodd}
\affiliation{Department of Astronomy and Astrophysics,
University of California, 
Santa Cruz, CA,  95064, USA }
\affiliation{DARK, Niels Bohr Institute,
University of Copenhagen, 
 Blegdamsvej 17, DK-2100 Copenhagen, Denmark}

\author[0000-0001-8825-4790]{Jamie A.P. Law-Smith}
\affiliation{Department of Astronomy and Astrophysics,
University of California,
Santa Cruz, CA,  95064, USA }
\affiliation{DARK, Niels Bohr Institute,
University of Copenhagen, 
 Blegdamsvej 17, DK-2100 Copenhagen, Denmark}
 
\author[0000-0002-4449-9152]{Katie Auchettl}
\affiliation{School of Physics, 
 The University of Melbourne, 
 Parkville, VIC 3010, Australia}
\affiliation{ARC Centre of Excellence 
 for All Sky Astrophysics in 3 Dimensions 
 (ASTRO 3D)}
 \affiliation{Department of Astronomy and Astrophysics,
University of California,
Santa Cruz, CA,  95064, USA }

\author[0000-0003-2558-3102]{Enrico Ramirez-Ruiz}
\affiliation{Department of Astronomy and Astrophysics,
University of California,
Santa Cruz, CA,  95064, USA }
\affiliation{DARK, Niels Bohr Institute,
University of Copenhagen, 
 Blegdamsvej 17, DK-2100 Copenhagen, Denmark}
 
\author{Ryan J. Foley}
\affiliation{Department of Astronomy and Astrophysics,
University of California,
Santa Cruz, CA,  95064, USA }

\defcitealias{2017ApJ...850...22L}{LS17}

\begin{abstract}
We study the properties of the host galaxies of Changing-Look Active Galactic Nuclei (CL AGNs) with the aim of understanding the conditions responsible for triggering CL activity.  We find that CL AGN hosts  primarily reside in the so-called green valley that is located between spiral-like star-forming galaxies and dead ellipticals, implying that CL AGNs are activated during distinct periods of quenching and galaxy transformation processes. CL AGN  hosts have low galaxy asymmetry indicators, suggesting that secular evolutionary processes (the influence of bars and spirals, and possibly minor mergers) might be the primary mechanism for transporting gas to the vicinity of the supermassive black hole (SMBH) rather than major mergers. Similar to tidal disruption events (TDEs) and highly variable AGNs, we find that CL AGN hosts are associated with SMBHs residing in high density pseudo-bulges and appear to overlap most significantly with the population of low-ionization nuclear emission-line region (LINER) galaxies. As such, CL AGN are likely 
fueled by strong episodic bursts of accretion activity, which appear to take place preferentially as the amount of material accessible for star formation and accretion dwindles.   We also identify that CL AGN hosts are
characterized by either large S\'ersic indices or high bulge fractions, which suggests a simple metric for
identifying candidates for spectroscopic follow-up observations in forthcoming synoptic surveys.
\end{abstract}


\section{Introduction} \label{sec:intro}
The nuclei of some galaxies undergo extreme activity, when the central super-massive black hole (SMBH) accretes gas at a sufficiently high level to be identified as an active galactic nucleus (AGN). Such activity is short-lived when compared to galactic lifetimes, and was common when the Universe was only about one fifth of its present age \citep{2001AJ....121...54F}. The growth of a SMBH is the imminent conclusion of such activity, and SMBHs that are now starved of fuel and therefore have switched off or have dimmed are significantly more abundant than active galactic nuclei (AGNs) in nearby galaxies \citep{2008ARA&A..46..475H}.

AGN activity in the local Universe appears to occur without the need for vigorous tidal encounters or major mergers \citep[e.g.,][]{2014ARA&A..52..589H}. Yet, significant changes in the mass supply to the SMBH, likely driven by non-axisymmetric perturbations, allow angular momentum to be efficiently transported from the gas to the dark matter or the stars\footnote{Illustrations of such  perturbations include, for example, bars, tidal tails and spiral arms \citep[e.g.,][]{2004ARA&A..42..603K,2008MNRAS.390L..69A,2014RvMP...86....1S}.}, and is observed to take place in nearby galaxies. Here gas can be accreted by the SMBH at rates sufficiently high enough to fuel Seyfert-like AGN activity, with changing-look (CL) AGN being one of the most extreme instances of such phenomenon \citep[][]{2004MNRAS.354..892M,2017ApJ...846L...7S, 2018ApJ...864...27S,2019ApJ...876...75C}. 

CL AGNs are sources that can exhibit drastic transitions between AGN types (transitions between type 1 and type 2 or vice-versa) and are characterized by the emergence and/or disappearance of broad emission lines. This transformation in emission features is accompanied by changes in the power-law optical/UV continuum on timescales of months to years that are much stronger than those seen in AGN \citep[e.g.,][]{1984PAZh...10..803L,2015ApJ...800..144L,2016MNRAS.457..389M,2016ApJ...826..188R,2016MNRAS.455.1691R,2017ApJ...835..144G,2018ApJ...862..109Y,2019MNRAS.486..123R,2019ApJ...883...94T,2019ApJ...883L..44G,2020ApJ...905...52G,2020MNRAS.tmp.2634W}. While it is generally accepted that some type of instability in the accretion disk may be responsible for CL activity \citep[e.g.][]{2015ApJ...800..144L}, in contrast  to, e.g., obscuration \citep[e.g.,][]{1989ApJ...340..190G,1992AJ....104.2072T,2017ApJ...846L...7S, 2018ApJ...864...27S}
or the  tidal disruption and subsequent digestion of a star \citep[e.g.,][]{2015MNRAS.452...69M,2019ApJ...883...94T,2020ApJ...898L...1R}, the  exact transport  mechanism has yet to be clearly identified as the radial transport of gas driven by viscous processes in a standard disk fails to explain the observed CL properties  \citep{2018NatAs...2..102L}. These highly variable sources, mostly uncovered serendipitously in nearby AGNs, can now be systematically discovered by  wide-field spectral and photometric searches and as of writing there are now several tens
of known CL AGN \citep[e.g.,][]{1986ApJ...311..135C,1993ApJ...410L..11S,2001ApJ...554..240E,2014ApJ...796..134D,2014ApJ...788...48S,2016MNRAS.457..389M,2018ApJ...862..109Y,2020ApJ...889...46S}. This recent observational progress, combined with the uniformity and wide sky coverage of the Sloan Digital Sky Survey \citep[SDSS;][]{2000AJ....120.1579Y,1998AJ....116.3040G,2006AJ....131.2332G}, provides us with the unique opportunity to study CL AGNs and their host galaxies in the context of carefully selected samples of nearby AGN host galaxies. The characterization of the properties of the host galaxies of CL AGN allows us to probe the link between the large-scale galaxy structure of these hosts and the occurrence of CL AGNs.

In this {\it Letter} we study the properties of CL AGN host galaxies in the context of a catalog of $\approx 5\times 10^5$ galaxies from the SDSS following the framework developed by \citet{2017ApJ...850...22L} (which we refer to in what follows as \citetalias{2017ApJ...850...22L}), which was previously used to analyze the host galaxies of tidal disruption events (TDEs). 
To this end,  we compare a sample of 17 CL AGN hosts that are found within our catalog and compare the properties of these hosts to the properties of a set of control samples of AGNs and galaxies in the local Universe. We explore a wide range of primary galaxy properties, including redshift, stellar mass, SMBH mass, star formation rate (SFR), bulge and surface brightness properties, S\'ersic index, bulge-to-total-light ratio, and galaxy asymmetry. We also compare the host galaxies of CL AGNs to the hosts of TDEs and highly variable AGNs across these observable qualities to better understand the properties of these extreme phenomena. 

The {\it Letter} is organized as follows. We describe our CL AGN sample as well as the galaxy catalog in Section~\ref{sec:methods}. The variability properties of our selected CL AGN sample are studied in detail in Appendix~\ref{sec:appendix_a}. In Section~\ref{sec:hosts}, we compare the properties of CL AGN host galaxies with  control samples of galaxies and AGN hosts in the local Universe. We discuss the CL AGN sample in the context of the proposed AGNs evolutionary sequence studied by \citet{2009ApJ...705.1336C} in Section~\ref{sec:evolution}. We contrast the properties of the CL AGN sample with those of TDEs, highly variable AGNs and passive galaxies in Section~\ref{sec:passive}. We summarize our findings in Section~\ref{sec:discussion}. 
\section{Galaxy Catalog and CL AGN Sample} \label{sec:methods}

\subsection{Reference Catalog} \label{subsec:catalog}
We use the reference catalog of \citetalias{2017ApJ...850...22L}. The catalog is comprised of bulge and disk decompositions and photometry from 
\citet{2011ApJS..196...11S}; bulge, disk, and stellar mass estimates from \citet{2014ApJS..210....3M}; and star formation rate (SFR) and spectral 
properties from the Sloan Digital Sky Survey MPA-JHU DR7 catalog\footnote{http://wwwmpa.mpa-garching.mpg.de/SDSS/DR7}
\citep{2004MNRAS.351.1151B}. After performing quality control measures as described in Section 2.1 of \citetalias{2017ApJ...850...22L}, the 
intersection of these three catalogs yields $\approx5\times10^5$ galaxies and makes up our reference catalog. 

The combination of the aforementioned sources yields a wide variety of galaxy properties that can be compared with those of our CL AGN host galaxy catalog. The \citet{2011ApJS..196...11S} study provides redshift, bulge $g-r$, bulge and galaxy magnitudes, galaxy half-light radius, galaxy S\'ersic 
index, bulge to total light fraction, galaxy asymmetry indicator as well as galaxy inclination. From  the work of \citet{2014ApJS..210....3M} we obtain bulge and total
stellar masses, and finally from the MPA-JHU catalog we have velocity dispersion, H$\alpha$ equivalent width, Lick H$\delta_{A}$, $\textrm D_{n}4000$ which is a indicator of the age of the galaxy stellar population, and SFR. We also estimate SMBH masses following the $M_{\rm bh}$--$\sigma_{e}$ scaling relation from 
\citet{2013ARA&A..51..511K}. The reader may refer to \citetalias{2017ApJ...850...22L} for a detailed discussion of this choice of estimation as well as for comparisons with other works. 

With the goal of comparing them to our sample of CL AGN, we define a sample of AGNs from our reference catalog using the BPT diagram designation \citep{2003MNRAS.346.1055K, 2006MNRAS.372..961K}, where the ratios of the OIII (5007\AA), $\textrm H\beta$, NII (6584\AA), and $\textrm H\alpha$ emission lines are used to classify them by requiring that the signal-to-noise for each of the lines used in this diagnostic be $\geq 3$. The total number of AGNs in the catalog is 93,304, while the number of high signal-to-noise AGNs is 52,613 \citepalias{2017ApJ...850...22L}.

\subsection{CL AGN Host Galaxies} \label{subsec:clagn_hosts}
To create our sample of CL AGNs, we compile a list of currently known (and published) CL AGNs or CL quasars and cross-match it with our reference catalog of galaxies. From this list we
find a total of 17 matches: 11 from \citet{2018ApJ...862..109Y}, 4 from \citet{2019ApJ...883...31F}, and 2 from \citet{2020ApJ...889...46S}. All 
but the last 2 are of the {\it turn-on} variety, meaning that the AGN luminosity increased by $\gtrsim$0.5 magnitudes between observations. 
In Table \ref{tab:info} in Appendix~\ref{sec:appendix_a}, we list our CL AGN sample and some of their key measurements. We note that the two CL AGN that were discovered while {\it turning-off}  have  S\'ersic index measurements  in our catalog but not  bulge-disk decompositions or stellar mass estimates. As such they are only included in S\'ersic index plots presented in this study.
\begin{figure*}
\epsscale{1.2}
\plotone{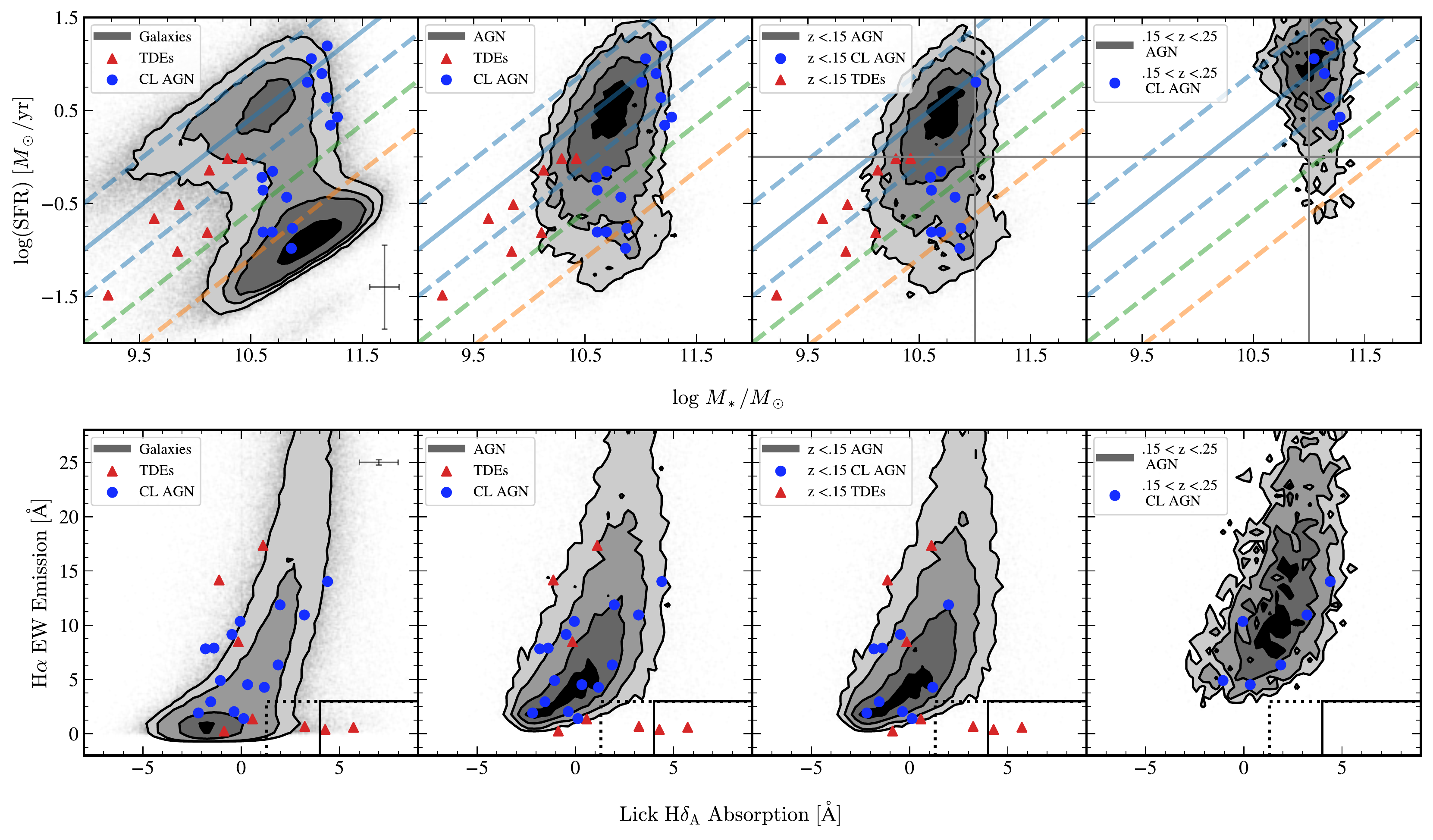}
\caption{\textit{Top panel:} Star formation rate (SFR) versus total stellar mass ($M_\ast$) for CL AGN host galaxies and TDE host galaxies from \citetalias{2017ApJ...850...22L}. All of the galaxies hosting CL AGNs  were identified as AGN in our catalog. Contours from left to right: galaxies, all AGNs, AGNs with redshift below 0.15, and AGNs with redshift above 0.15. Median errors for the CL AGN host galaxy measurements
are shown in the bottom right of the leftmost panel. The solid blue line represents the star-forming main sequence \citep[SFMS;][]{2010ApJ...721..193P}. As in \citetalias{2017ApJ...850...22L}, dashed lines are spaced 1$\sigma$ apart, with the green valley falling 1-3$\sigma$ 
below the SFMS (the region between the lower blue dashed line and orange dashed line).
Note how the gray lines in panels 3 and 4 cleanly divide the two populations of CL AGNs in total stellar mass and SFR. \textit{Bottom panel:} H$\alpha$ 
equivalent width versus Lick H$\delta_A$ absorption for CL AGN and TDE hosts, following 
\citet{2016ApJ...818L..21F} and \citetalias{2017ApJ...850...22L}. Contours same as top panel. Median errors for the CL AGN host galaxy measurements are shown in the top right of leftmost 
panel. Galaxies inside the solid dashed lines constitute our selection of E+A galaxies, as defined by 
\citet{2016ApJ...818L..21F} and \citetalias{2017ApJ...850...22L}, while galaxies that fall in the area between the dotted black lines designate quiescent Balmer-strong galaxies.
\label{fig:fig1}}
\end{figure*}

Given the different selection functions used to uncover CL AGNs \citep{2020ApJ...889...46S}, we do not expect to arrive at an unbiased representative sample of CL AGNs using the method outlined above. For example, 15 out of 17 of our CL AGNs are of the {\it turn-on} variety. The cause of this over representation of {\it turn-on} CL-AGN is twofold. First, current detection methods are biased toward finding objects that {\it turn-on}, given the over representation of quiescent galaxies in the local Universe. Second, there is a clear redshift dependence with {\it turn-off} objects being discovered at higher redshifts than {\it turn-on} CL AGN and thus being underrepresented in our catalog. Further discussion of this under representation can be found in Appendix~\ref{sec:appendix_a}.

With this in mind, we analyze the properties of our \textit{turn-on} CL AGN host galaxies\footnote{We note that all of CL hosts  are categorized as AGNs in  our reference catalog using the BPT diagram designation.} by comparing them with the properties of the galaxies found in our reference catalog.
This is in contrast to the recent study  of \citet{2019ApJ...876...75C}, which analyzed the host galaxy properties of 4 {\it turn-off} CL quasars at $z> 0.198$ imaged by the Gemini Multi Object Spectrograph. We should note that our reference catalog, which is derived predominantly from SDSS, is intrinsically flux limited, implying that the number density of galaxies decreases with redshift.
As such, this limits our sample of galaxies and CL AGN to those within $z \lesssim 0.25$ (see Figure \ref{fig:figa1}). However, as the majority of our CL AGN sample (Table \ref{tab:info}) is found at $z \lesssim 0.17$, this should not significantly affect the results of our analysis. 
Nonetheless, to further minimize the possible effect of this on our analysis, we create control samples that have been matched on key galaxy properties. 

\section{The Galaxy Hosts of CL AGN\lowercase{s}} \label{sec:hosts}

\subsection{CL AGNs in the Context of Nearby Galaxies} \label{subsec:overview}
Star-forming galaxies appear to obey a clear relation between SFR and stellar mass \citep{2007ApJ...660L..43N, 2007A&A...468...33E, 2007ApJ...670..156D, 2011ApJ...739L..40R}, which is commonly referred to as the star formation main sequence (SFMS). Nearby galaxies along the main sequence (solid blue line in the top left panel in Figure~\ref{fig:fig1}) are disk-dominated and show low levels of dynamical turmoil \citep{2009ApJ...706.1364F, 2011ApJ...742...96W}, while starburst galaxies, characterized by having elevated SFR and residing above the main sequence, show structural properties consistent with being recently disturbed \citep{2011ApJ...742...96W, 2012ApJ...757...23K}.

Following \citetalias{2017ApJ...850...22L}, the 
dashed blue lines above and below the SF main sequence solid blue line in Figure~\ref{fig:fig1} are spaced by 1$\sigma$, corresponding to 0.5 dex (the median scatter of the SFR values). Over time, galaxies are seen to move from the concentration in the top left, corresponding to spiral galaxies, to the bottom right, where elliptical galaxies are located. Much is still debated about galaxies located within the transition region between spiral and elliptical galaxies. This region is commonly referred to as the {\it green valley} and lies between the lower dashed blue line and above the orange dashed line in Figure~\ref{fig:fig1} (which broadly denotes the transition to elliptical galaxies). 
In the upper panel of Figure~\ref{fig:fig1} moving from left to right, we include the distribution of all galaxies in our catalog, all galaxies which harbor AGNs based on the BPT diagram, lower-redshift AGNs ($z < 0.15$), and higher-redshift AGNs ($0.15 < z < 0.25$) as contours. Note all 2D contour plots\footnote{A reminder that for two-dimensional contour plots, 1$\sigma$ contains 39.3\% of the population and 2$\sigma$ contains 86.5\%.} presented in this work have contours which are spaced by 0.5$\sigma$.

While there is an open debate about the relative roles of SFR and total stellar mass ($M_\ast$) on the luminosity function of AGNs \citep[e.g.][]{2019MNRAS.484.4360A}, 8/15 of our CL AGN sample are located within the green valley region of this diagram, and the majority of the sample lies within 1 sigma of this designation, as seen in the top panels in Figure~\ref{fig:fig1}. \citet{2019MNRAS.484.4360A} find that the AGN fraction is enhanced for galaxies located below the SF main sequence (blue solid line) compared with those on the main sequence. Indeed, we find that although AGN comprise about 10\% of our sample, we find that 15\% of galaxies located in the green valley host an AGN. As discussed in greater detail in \citet{2004MNRAS.351.1151B}, SFR for AGN in our sample are estimated using the strength of the 4000 angstrom break because the line fluxes are likely to be affected by the AGN component. We caution that uncertainties in this relationship are of the order 0.5 dex as illustrated  by the associated error estimate plotted in Figure~\ref{fig:fig1}.

The preferential location of the majority of CL AGN hosts between these two families of galaxies, even when compared to AGN hosts, seems to imply that these sources are more effectively activated in particular periods of the quenching \citep[e.g.,][]{2020ApJ...899L...9G} and galaxy transformation \citep[e.g.,][]{2019ApJ...876...75C} processes, which is bolstered by the fact that the majority of galaxies in the green valley do not host AGNs (see Figure \ref{fig:fig1} upper left panel). 

We find that the CL AGN hosts appear to lie in two fairly distinct groups (as marked by the gray solid lines in the upper right two panels of Figure \ref{fig:fig1}), which becomes more obvious after dividing the samples into higher ($0.15 < z < 0.25$) and lower ($z <0.15$) redshift ranges.  

TDEs also display similar flaring signatures as CL AGNs \citep{2018ApJ...852...37A,2019ApJ...883...94T,2020ApJ...898L...1R}; thus, it is useful to compare the host galaxy properties of the eight TDEs from \citetalias{2017ApJ...850...22L} which have measurements from the MPA-JHU catalog, which are plotted as red triangles in Figures \ref{fig:fig1} and \ref{fig:fig5}, to those of our CL AGN sample.  We find that similar to the TDE population (\citetalias{2017ApJ...850...22L}), nearly all of our low redshift CL AGN sample falls within the green valley, while the majority of AGNs found at similar redshifts are consistent with the SFMS \citep[][]{2020arXiv200101409V,2020arXiv201010738H}. In contrast, the higher redshift population of CL AGN are more consistent with the AGN population of similar redshifts. It is, however,  worth noting that the number of galaxies in our catalog that fall within the green valley decreases significantly at the higher ($0.15 < z < 0.25$) redshift range. Additional discussion of the potential bias against observing CL AGNs in green valley galaxies can be found in Appendix~\ref{sec:appendix_b}.

\begin{figure*}
\epsscale{1.15}
\plotone{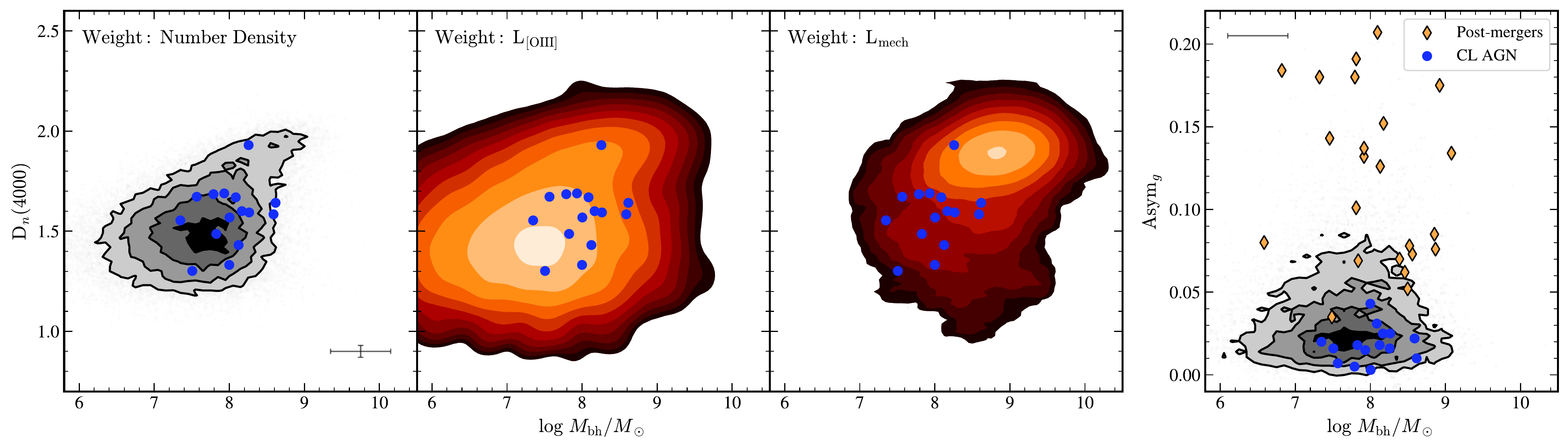}
\caption{\textit{Left three panels:} Strength of the 4000 angstrom break versus black hole mass. The far left contours are AGN from our sample catalog, while the following two are SDSS galaxies weighted by [OIII] line luminosity (corresponding to bolometric radiative luminosity; middle panel) and jet mechanical luminosity (right panel) adapted from \citet{2014ARA&A..52..589H}. Median errors for the CL AGN host galaxy measurements are shown in the bottom right of the leftmost panel. \textit{Right panel:} Asymmetry ($g$-band) versus black hole mass for CL AGN hosts and our post-merger sample \citep{2013MNRAS.435.3627E}. Contours are AGN matched in redshift to our CL AGN sample using a 7.5\% tolerance and capped at $2\times10^3$ matches per CL AGN. Median error in CL AGN SMBH mass is shown in the top left (there is no significant error associated with asymmetry measurements in our catalog). \label{fig:fig2}}
\end{figure*}

Further details about a galaxy's star forming state can be determined by considering H$\alpha$ equivalent width (EW) emission and H$\delta$ absorption (Figure \ref{fig:fig1} lower panels). Higher values of the former correspond to increased current star formation, while higher values of the latter suggest a large population of A-type stars, whose atmospheres absorb strongly at this wavelength regime. Objects with low H$\alpha$ emission and high H$\delta_A$ absorption thus 
represent galaxies that have experienced a starburst around the time when the A stars become prominent, which is $\approx0.1-1$ Gyr ago. These are commonly referred to as E+A-type galaxies \citep{1996ApJ...466..104Z,2020SSRv..216...32F}.
\citetalias{2017ApJ...850...22L} define an E+A sample by requiring H$\delta_{A}-\sigma$H$(\delta_{A}) > \ 4.0$ and H$\alpha$ EW  $< \ 3.0$ (where $\sigma$H$(\delta_{A})$ is the error in the Lick  H$(\delta_{A})$ index). This designation is represented by solid black lines in each of the bottom panels of Figure \ref{fig:fig1}. We find that CL AGN hosts follow the underlying distribution of AGN hosts in this SF plane, distinct from a majority of TDEs, which have a strong preference for being detected in E+A-type galaxies \citep[e.g.,][]{2020SSRv..216...32F}. We note that for AGN, H$\alpha$ EW is only an approximation for SFR given the potential for contamination from the central source. Having said this, we don't expect more than 40\%  of the H$\alpha$ flux to have a non-stellar origin for any galaxy in the sample \citep{2004MNRAS.351.1151B}.

\subsection{Radiative-Mode AGNs and Gas Fueling} \label{subsec:agn_mode}
The incidence of optical nuclear activity depends strongly on the $D_n (4000)$ line index \citep{2014ARA&A..52..589H}, which is broadly indicative of  stellar age. In the left panel of Figure~\ref{fig:fig2}, we plot our $D_n (4000)$ as a function of $M_{\rm bh}$ for our reference AGN sample as the black-gray contours, while we mark our CL AGN sample as the blue dots. When compared to galaxies in the $D_n (4000)- M_{\rm bh}$ plane, AGNs have an incidence fraction that is at least an order of magnitude larger in the youngest post-starburst galaxies \citep{2014ARA&A..52..589H}, with our CL AGN sample having a $D_n (4000)$ consistent with the overall AGN population. This overabundance is likely linked with the noticeable increase in molecular gas availability toward younger ages, which is expected to lead to efficient fueling \citep{2020ApJ...899L...9G}.  

\begin{figure*}
\epsscale{1.1}
\plotone{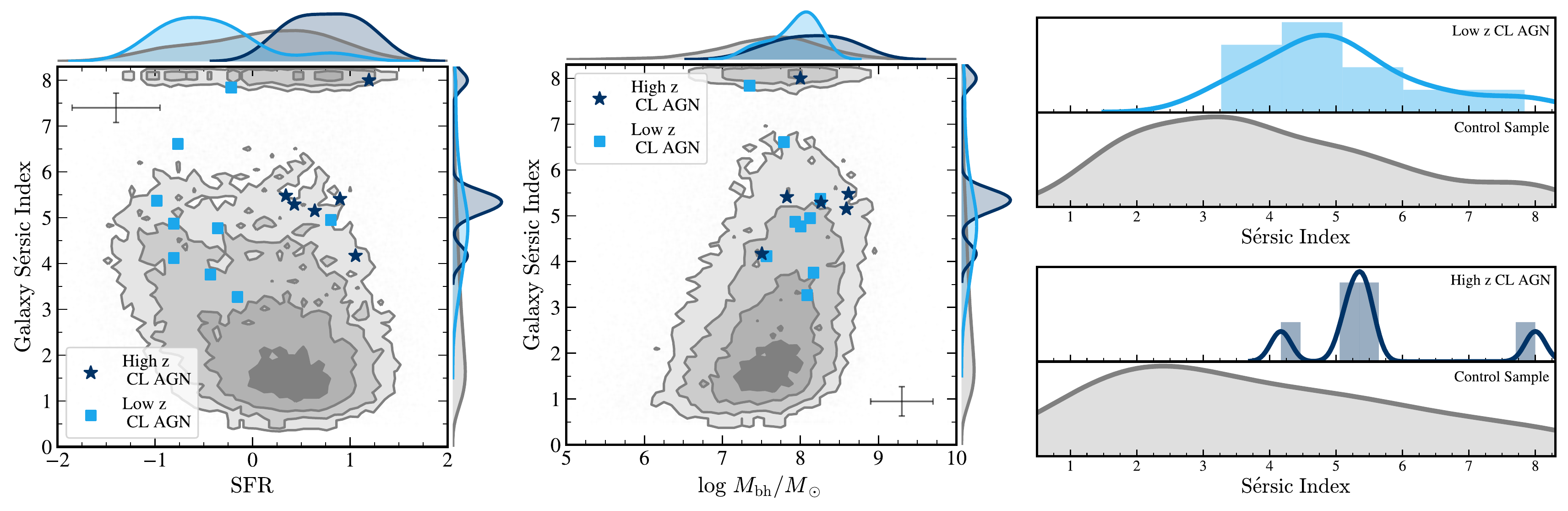}
\caption{\textit{Left panel:} S\'ersic index versus star formation rate (SFR) for CL AGN hosts split by redshift, with low $z$ CL AGN hosts ranging from $z <
0.15$ and high $z$ CL AGN hosts ranging from $0.15 <  z < 0.25$. Median errors for CL AGN hosts are displayed in the top left. 1D projections are 
smoothed and normalized to have equal probability area. \textit{Middle panel:} S\'ersic index versus black hole mass for the same redshift groups of CL AGN hosts. Median errors  are shown in the bottom right, and 1D projections are again smoothed and normalized. 
\textit{Right panel:} Histograms of S\'ersic index for low $z$ CL AGNs (top) and high $z$ CL AGNs (bottom) with their respective control sample catalogs using the method developed by \citetalias{2017ApJ...850...22L}, which are matched on  SFR, total stellar mass, and redshift to a 7.5\% tolerance. We note that this tolerance is larger than the one used by \citetalias{2017ApJ...850...22L} given the smaller number of AGNs as compared to regular galaxies.  \label{fig:fig3}}
\end{figure*}

The local population of AGNs can be broadly divided into two distinct groups \citep{2014ARA&A..52..589H}.
The first group consists of objects whose dominant power is in the form of radiation from accretion at rates in excess of $\approx$1\% of the Eddington Limit and are commonly referred to as radiative-mode AGNs. These objects are preferentially found in lower mass galaxies with a slight preference for younger stellar populations (middle panel in Figure~\ref{fig:fig2}). The second group is comprised of objects whose primary power is from mechanical energy produced from outflows rather than accretion. AGNs in this group reside in the most massive galaxies and generally possess old stellar populations (middle panel in Figure~\ref{fig:fig2}). Not surprisingly, CL AGN hosts closely follow the distribution of the radiative-mode AGN sample. As these galaxies are found to have disks of cold gas near the nucleus \citep{2020ApJ...899L...9G}, this suggests fuel is more readily available to undergo episodes of sporadic and bright flaring.

The activation of SMBHs in the local Universe appears to occur without the need for major mergers \citep{2014ARA&A..52..589H} and is largely fueled by secular processes\footnote{Here secular processes correspond to the slow rearrangement of energy and mass that results from interactions of bars, spiral structure, and dark halos \citep{2004ARA&A..42..603K}.} that can transport inward the abundant supply of cold, dense gas in the central regions of these galaxies. Visual inspection of [O III]-selected AGN hosts on radial scales on the order of a kiloparsec commonly reveals rings, spiral arms, bars and dust lanes \citep[e.g.][]{1998ApJS..117...25M,2014ARA&A..52..589H}, suggesting that this process is dominant in these galaxies. The general idea that AGN activation takes place primarily in galaxies that show high stellar mass concentrations in the central regions and with larger disordered motions is reinforced by the MaNGA survey study recently reported by \citet{2018RMxAA..54..217S}.  

Motivated by this, we analyze the asymmetry measurements of our sample of CL AGN hosts. Here we compare the asymmetries of these hosts with a sample of all 27 post-merger galaxies belonging to our catalog that exhibit enhanced star formation rates and increased incidences of AGN as identified from SDSS \citep{2013MNRAS.435.3627E}. Results are plotted in the right panel of Figure \ref{fig:fig2}. Interestingly, we find that our CL AGN host sample has small asymmetry indicators compared to our post-merger sample, suggesting that they are not the products of major mergers. Furthermore, we find that CL AGN hosts have a narrower distribution in asymmetry indicators when compared to our reference AGN catalog. This suggests CL AGN have even higher central stellar concentrations/larger disordered motions compared to that seen in the general AGN population.  Interestingly, CL AGN and TDEs show similar asymmetry indicators (see Figure 6 middle panel of \citetalias{2017ApJ...850...22L}). We caution, however, that given the SDSS resolution limit a small asymmetry value does not necessarily imply a lack of a merger or a tidal encounter. This is because these indicators tend to be only large for major mergers in gas-rich galaxies \citep{2010MNRAS.404..575L,2014A&A...566A..97J}. The reader is referred to Appendix~\ref{sec:appendix_d} for further discussion on this issue.

While we find that major mergers are not the dominant fueling mechanism for the vast majority of our CL AGN sample based on their low asymmetry measurements, the idea of a causal connection between 
tidal interactions and the ignition of the CL AGN activity has been demonstrated by \citet{2019ApJ...876...75C}. By utilizing Gemini imaging, which is capable of detecting fainter features than typical SDSS imaging, \citet{2019ApJ...876...75C} conclude that three of their four CL quasar host galaxies ($z\gtrsim 0.2$), all of them belonging to the {\it turn-off} category, show clear evidence of extended tidal tails or tidal streams on radial scales on the order of tens of kiloparsecs from the nucleus. If activation in \textit{turn-on} CL AGN is largely fueled by non-axisymmetric perturbations in the underlying mass distribution, we would therefore expect CL AGNs to be found primarily in host galaxies that have higher bulge fractions than “typical” AGN galaxies at similar SMBH masses. It is to this question that we now turn our attention. 

\subsection{S\'ersic Index and Bulge-to-total-light Ratio} \label{subsec:sersic}

The S\'ersic index is a measure of the steepness of the light profile in galaxies and is generally assumed to effectively trace the central concentration of 
stellar light and to a lesser extent the central stellar density. In Figure~\ref{fig:fig3} we show S\'ersic index versus SFR (left panel) and S\'ersic index versus $M_{\rm bh}$ (middle panel) for our  CL AGN sample and our reference AGN catalog. Similar to what was done in Figure~\ref{fig:fig1}, we split our CL AGN sample by redshift
to better control for the S\'ersic index dependence with redshift \citepalias{2017ApJ...850...22L}. We find that all of the CL AGN host galaxies have high S\'ersic indices for their corresponding SMBH masses compared to our AGN sample, regardless of redshift. In the right panel of Figure~\ref{fig:fig3}, we create two samples of AGN matched to the high and low $z$ CL AGN groups based on SFR, total stellar mass, and redshift and compare their distributions of S\'ersic indices. We find that CL AGN hosts have distributions of S\'ersic indices that peak systematically at much higher values compare to their AGN matched samples. We also note that CL AGN tend to have higher S\'ersic indices (S\'ersic indices $\approx 4-5$) compared to TDE hosts which show a range (S\'ersic indices $\approx 3$) \citepalias{2017ApJ...850...22L}.

The relatively high central concentration of light in CL AGN hosts is also observed in their bulge-to-total-light ratio distributions, however due to the similarity with that seen in Figure \ref{fig:fig3} we do not include this figure. Values of bulge-to-total-light ratios can be found in Table \ref{tab:info}. Interestingly, as found by \citetalias{2017ApJ...850...22L} for TDE hosts, we have identified a photometric criterion (S\'ersic index or bulge-to-total-light ratio) that could help predict the occurrence of CL behavior in a fraction of AGN hosts. 

From the above, our results imply that the activation of CL AGNs in the local Universe primarily occurs in centers of galaxies where two criteria are met: there is an abundant supply of cold gas (whose presence is implied by a relatively recent starburst episode or ongoing star formation) and there is a large concentration of stellar mass in the central regions. The mechanism by which gas is then subsequently transported to the vicinity of the SMBH remains an open question. Yet, perturbations in the underlying centrally concentrated  mass distribution are expected to cause strong episodic feeding events, presumably related to the short dynamical times near the nucleus.

\section{CL AGN\lowercase{s} in The Evolutionary Context} \label{sec:evolution}

\begin{figure*}
\epsscale{0.95}
\plotone{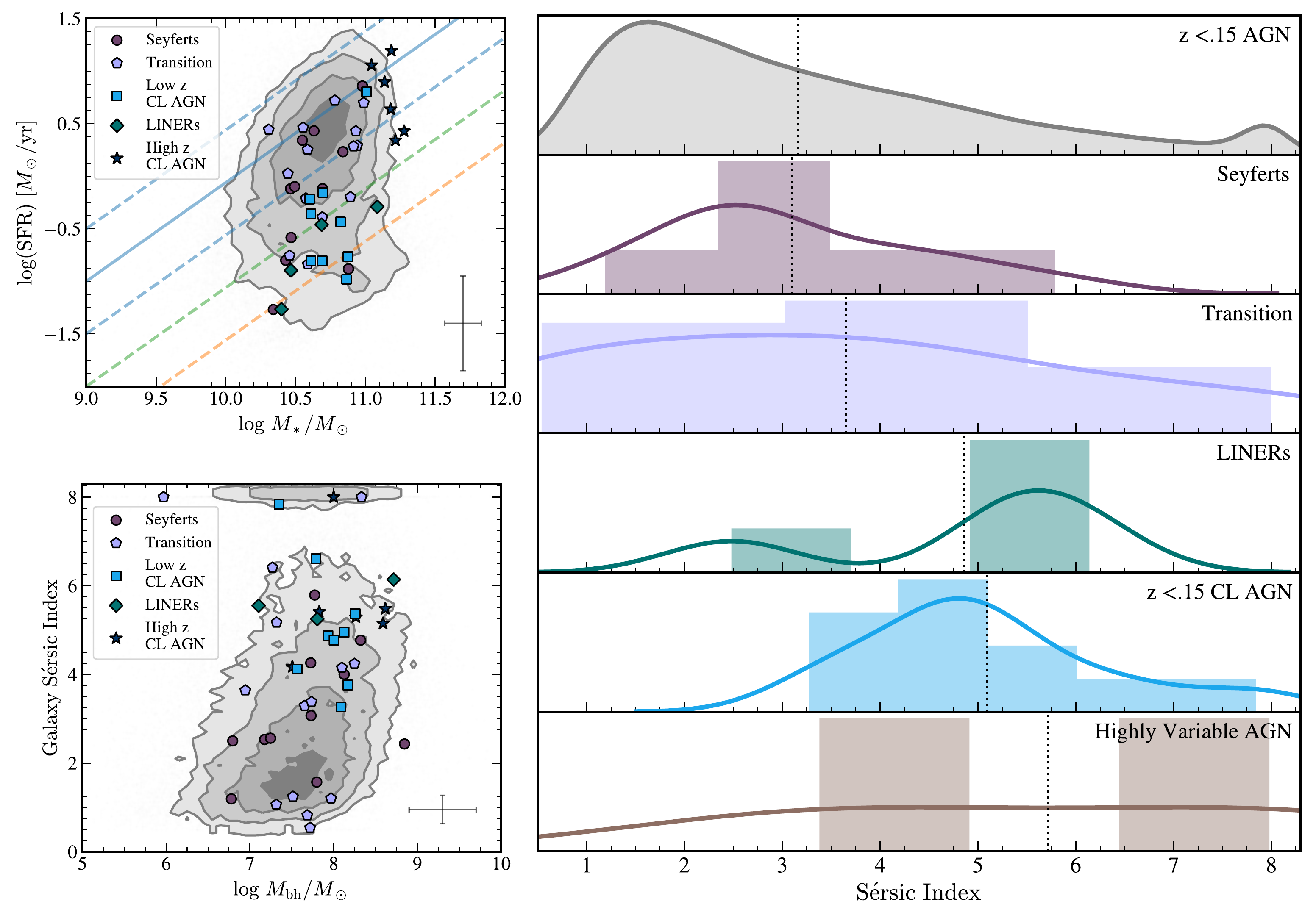}
\caption{\textit{Upper left panel:} SFR versus total stellar mass, as in Figure \ref{fig:fig1}. This time we plot the low $z$ CL AGN sample along with a similar $z$ range mass-limited selection of Seyfert, Transition, and LINER host galaxies from \citet{2009ApJ...705.1336C}. Low $z$ AGNs are shown as contours, each shade spaced by 0.5$\sigma$. For completeness we also include the high $z$ CL AGN sample. The solid blue line again represents the SFMS and dashed lines are spaced by 1$\sigma$, with the green valley falling between the lower blue dashed line and orange dashed line. Median errors for the low $z$ CL AGN hosts are shown in the bottom right. 
\textit{Lower left panel:} Galaxy S\'ersic index versus log SMBH mass for the same groups.
Median errors for CL AGN hosts are again found in the lower right. \textit{Right:} Histograms of S\'ersic index for the samples described 
previously (with the exception of the high $z$ CL AGN sample, which has a different redshift range.), with the addition of four highly variable AGNs for comparison: KUG 1624+351 \citep{2014MNRAS.444.1041K}, J094608+351222 
\citep{2017MNRAS.470.4112G}, 2MASS J09392289+3709438 \citep{2016A&A...592A..74S}, and Swift J1200.8+0650 
\citep{2007ApJ...669..109L}. Each histogram is normalized to equal area, and solid lines represent the smoothed distribution. Dotted black lines indicate
median values of S\'ersic index for each population. \label{fig:fig4}}
\end{figure*}

As argued in Section \ref{subsec:overview}, optically selected AGNs are over-represented compared to galaxies in the so-called \textit{green valley} between spiral-like star-forming galaxies and ellipticals (Figure~\ref{fig:fig4}). The location of AGN hosts in this transition region seems to give credence to the connection between quenching and the activation of the SMBH \citep{2014ARA&A..52..589H}. Moreover, the fact that a large fraction of galaxies in the green valley do not host active SMBHs suggests that AGNs are activated in distinct instances during the quenching and galaxy transformation processes. 

SMBH activity in nearby galaxies is diverse, with AGN luminosities spanning many orders of magnitude in Eddington ratio \citep{2004ApJ...613..109H, 2008ARA&A..46..475H}. The most ubiquitous types of AGNs identified at optical wavelengths are either a LINER (Low-ionization Nuclear Emission-line Region) or a Transition object, whose properties fall somewhere in between starburst galaxies and AGNs \citep{2014ARA&A..52..589H}. Some LINERs exhibit quasar-like broad lines and are unambiguously accretion-powered while some lack these features and, as such, their emanating power could be generated by processes other than SMBH accretion such as shocks or radiation from hot stars \citep{2014ARA&A..52..589H}.

As galaxies quench and move off the SFMS, an evolutionary progression has been identified for the co-evolution of AGNs and their hosts despite the large variations within specific types:~HII~$\rightarrow$~Seyfert/Transition Object~$\rightarrow$~LINER~$\rightarrow$~Passive \citep{2006ApJ...650..727C, 2007MNRAS.382.1415S, 2008ApJ...673..715C, 2009ApJ...705.1336C}. This sequence  suggests a pathway for the transmutation of galaxies from SF (HII) via AGN to ultimately quiescence (Passive). The AGN sequence is thought to begin with Seyferts, which are considered to be the low-luminosity equivalent of quasars \citet{2009ApJ...705.1336C}. Broadly speaking, this sequence is expected to trace a decrease in SMBH accretion rate, as inferred by X-ray observations, and an aging progression of the stellar population in the nuclei.

We attempt to place CL AGNs within this framework in order to understand their activation properties. To do this, we use the sample compiled by \citet{2009ApJ...705.1336C}, which combined X-ray observations taken as a part of the Chandra Multiwavelength Project (ChaMP) with a sample of SDSS DR4 nearby galaxies, creating a sizable sample of AGNs that spans a range of AGN/galaxy types, from passive to actively line-emitting systems, as well as star-forming and accreting objects. Our results are shown in Figure \ref{fig:fig4}. Plotted is our low redshift CL AGN sample, described in Section \ref{subsec:sersic}, alongside the various members of the evolutionary sequence for which we have imposed stellar mass constraints by requiring that they fall within the minimum and maximum stellar mass of the CL AGN hosts. In addition, both sets of samples occupy a comparable redshift range. We also plot our higher $z$ CL AGN hosts for the sake of comparison but do not include them in the evolutionary sequence because of their difference in redshift range.

In the upper left panel of Figure \ref{fig:fig4}
we begin our comparison in the SFR versus total 
stellar mass plane. We include the same lines as described in Figure~\ref{fig:fig1}, with the SFMS as a solid blue line and the average location of the green valley denoted by the dashed green line. While the distribution of each AGN type exhibits considerable scatter in this plane, the proposed general trend of decreasing SFR along the evolutionary sequence appears to hold. Our sample of CL AGNs are clearly in the process of quenching and overlap most significantly with the Transition and LINER populations \citep[the reader is referred to][for further discussion of CL LINERs]{2019ApJ...883...31F}.  

In the lower left panel of Figure \ref{fig:fig4} we add additional information regarding the central concentration of light for the various host samples, which is done here by comparing their S\'ersic indices\footnote{We note the following trends are also apparent when using their bulge-to-total-light ratios.}. In the SFR versus total stellar mass case, we see considerable spread amongst each subgroup, yet we see median values that agree with the general trend of hosts having more compact and denser central regions as quenching moves forward. This is also apparent in the distributions of S\'ersic indices plotted in the right panel of Figure \ref{fig:fig4}. CL AGNs show a strong preference for residing in centrally concentrated, bulge-dominated galaxies, clearly placing them near the hosts of LINERs in the proposed evolutionary sequence. 

LINERS are also found to have, on average, the lowest X-ray luminosities for their corresponding SMBH masses \citep{2009ApJ...705.1336C}. We note that we found no clear correlation between the observed X-ray luminosity (relative to the Eddington limit) and the S\'ersic index of the SMBH hosts. This is expected given that the characteristic duration over which a galaxy responds and relaxes into a new state of dynamical equilibrium is much longer than the duty cycle of AGNs \citep{2002MNRAS.335..965Y}.

For completeness, we also include a selection of four highly variable AGNs that lie within the footprint of our reference catalog: KUG 1624+351 \citep{2014MNRAS.444.1041K}, J094608+351222  \citep{2017MNRAS.470.4112G}, 2MASS J09392289+3709438 \citep{2016A&A...592A..74S}, and Swift J1200.8+0650  \citep{2007ApJ...669..109L}. Their hosts are shown at the bottom of the S\'ersic histograms and have a similarly high median value as in the case of CL AGN hosts. 

These findings reinforce the idea that CL AGN activation and highly variable AGNs are preferentially found in green valley galaxies with high central stellar densities. Furthermore, strong episodic bursts of accretion activity appear to occur preferentially as the amount of material available for star formation and accretion dwindles. Our findings are consistent with the claim that lower luminosity AGN display greater variability amplitudes \citep{2014MNRAS.444.3078G}. 

\begin{figure*}
\epsscale{1.2}
\plotone{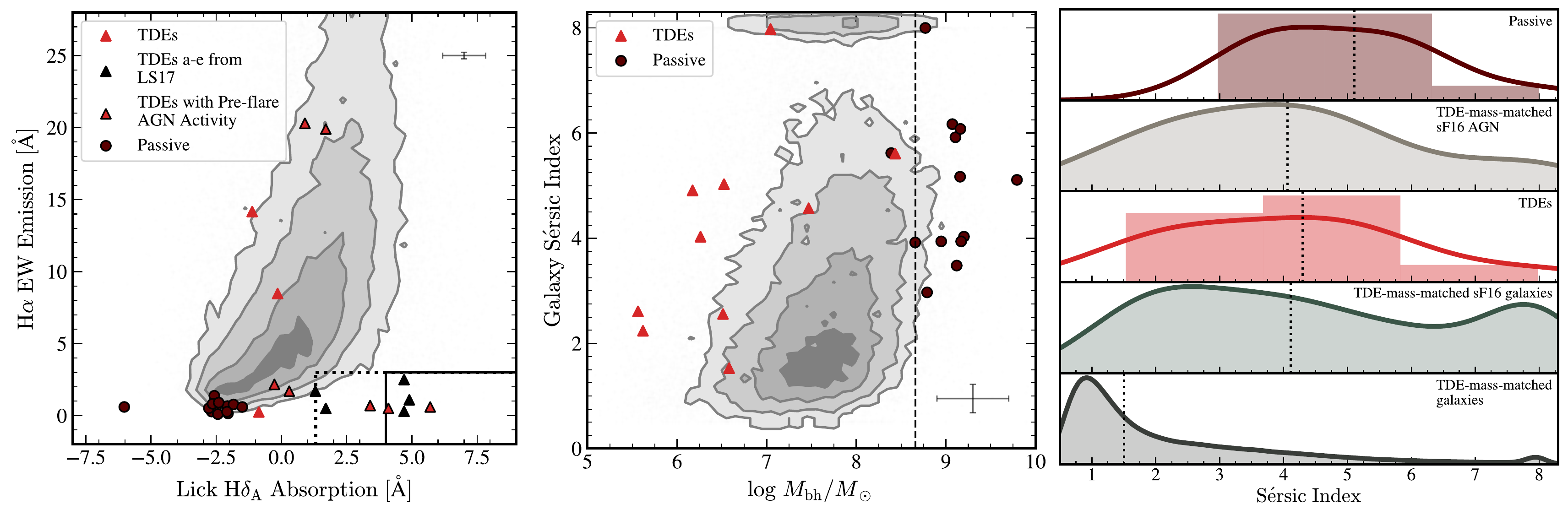}
\caption{\textit{Left panel:} H$\alpha$ equivalent width versus Lick H$\delta_A$ absorption for Passive galaxies and TDEs, with AGNs as contours. TDEs with measured pre-flare AGN activity \citep{2020SSRv..216...32F} are shown as red triangles outlined in black, and we additionally include TDEs a-e from \citetalias{2017ApJ...850...22L} as solid black triangles. As in Figure 
\ref{fig:fig1}, solid black lines designate the E+A selection region, while the area between the dotted black lines shows the quiescent Balmer-strong region. 
Median error in Passive host galaxy measurements is shown in the top right. \textit{Middle panel:} S\'ersic index versus black hole mass for the same populations. 
The dashed line represents the upper limit on SMBH for a TDE to be visible for a Sun-like star and a maximally spinning SMBH (a=0.999), as 
calculated by \citet{2012PhRvD..85b4037K}. At masses above this, tidal disruption would occur inside the event horizon. Median errors are shown in the
bottom right. \textit{Right panel:} Smoothed and area-normalized histograms of S\'ersic index. We compare the Passive galaxies and TDE hosts to a matched sample of  galaxies and AGNs in the E+A box (sF16 selection) as well as all galaxies, with the added constraint that they have SMBH masses within the  range of TDEs 1-5 (as defined in \citetalias{2017ApJ...850...22L}). Median values are shown by dotted black lines. \label{fig:fig5}}
\end{figure*}

\section{TDE\lowercase{s} and the Quenching of AGN\lowercase{s}}\label{sec:passive}
The relationship between the structure of a galaxy and the incidence and strength of its nuclear activity provides us with an opportunity to uncover the underlying AGN fueling mechanisms.
The evolution of AGNs within the galaxy transition region argues in favor of a causal connection between quenching and SMBH activity. The timescales of both mechanisms are, however, quite different. Estimates of the duty cycle of AGNs are $\approx 0.1$ Gyr \citep[e.g.,][]{2015A&A...583A..89S}, while the age differences for galaxies in the transition region indicate that quenching can last for a few Gyr \citep{2014ARA&A..52..589H}. This strongly suggests that AGNs are activated during brief periods of the transformation process and as such, it is not surprising that many green valley galaxies are inactive. 

Another category of galaxies considered by \citet{2009ApJ...705.1336C} that are believed to represent the tail end of the evolutionary AGN sequence are the so-called Passive galaxies, which are selected based on their lack of strong emission features in their spectra. In these Passive objects, the dense nuclear stellar 
environments (right panel in Figure \ref{fig:fig5} showing the relative high S\'ersic indices for this population) amplify the probability of a close stellar encounter with the central SMBH, and the lack of a strong SMBH accretion signature means that TDE flares can be more readily detected, if these galaxies have SMBH masses $\lesssim7\times10^8 M_{\odot}$ \citet{2012PhRvD..85b4037K}. 

It would thus be instructive to study the expectation of TDE activation within the evolutionary sequence and in particular in  these Passive galaxies. In Figure \ref{fig:fig5} we plot H$\alpha$ equivalent width  emission and H$\delta$  absorption for Passive galaxies, TDE hosts and the general AGN population. Passive galaxies identified by \citet{2009ApJ...705.1336C}  harbor SMBH that are too massive to disrupt Sun-like stars (middle panel in  Figure \ref{fig:fig5}), have little ongoing SF, and have stellar populations that are older than the $\lesssim$ 1 Gyr age associated with E+A galaxies (left panel in  Figure \ref{fig:fig5}). 

Within this evolutionary framework, it is thus surprising  that TDEs are found to preferentially occur in E+A galaxies, given their scarcity. The Balmer absorption in these hosts points to a sizable starburst population of approximate age $\approx$ 0.1-1 Gyr, while low H$\alpha$ indicates a lack of current star formation as also seen in the Passive galaxy sample. These E+A galaxies are rare: they comprise only $\approx$ 2\% percent of the local population, yet host more than half of the optically-detected TDE candidates to date \citep[e.g.,][]{2020SSRv..216...32F,2020arXiv201010738H}. 
This suggests that the TDE rate is likely enhanced by a dynamical mechanism that operates efficiently for $\approx$ 0.1-1 Gyr after the starburst episode \citepalias{2017ApJ...850...22L}.
Interestingly, even though $\approx 0.6\%$ of AGN hosts show E+A signatures \citepalias{2017ApJ...850...22L}, a non-negligible fraction of E+A TDE hosts display  pre-flare AGN activity \citep[][left panel in  Figure \ref{fig:fig5}]{2020SSRv..216...32F}. This suggests that TDE rates could also be potentially enhanced by the presence of a pre-existing accretion disk. It is therefore not surprising that TDEs have been invoked to trigger CL-like activity \citep{2020arXiv200307365P, 2019ApJ...883...94T}. CL AGNs and TDEs are therefore not two clearly separated populations within the evolutionary sequence. Yet, irrespective of their connection, it is evident that their activation preferentially occurs in a particular set of galaxies, those that show larger concentrations of stellar mass in the central regions (right panel in Figure \ref{fig:fig5}). 

\section{Summary and Conclusions}\label{sec:discussion}
We study the properties of CL AGN host galaxies in the context of a catalog of $\approx 5 \times 10^5$ galaxies in the local Universe. Our main conclusions, with the stated limitation that our sample includes primarily {\it turn-on} CL AGNs, are:
\begin{itemize}

\item CL AGN hosts are mainly located in the so-called \textit{green valley} region between spiral-like star-forming galaxies and dead ellipticals, even  when  compared  to our reference AGN catalog. The origin of the fueling gas in these systems can then be attributed to a recent star-burst episode or to a modest level of star formation.

\item CL AGN hosts have low galaxy asymmetry indicators, suggesting that major mergers are not the primary mechanism for transporting this gas inward and that secular processes might instead lead to CL activation during specific instances of the AGN quenching and galaxy transformation process (Appendix~\ref{sec:appendix_d}).  

\item CL AGNs, highly variable AGNs and TDE hosts have high galaxy S\'ersic indices and high bulge-to-total-light ratios for their corresponding SMBH masses, implying high stellar density in their cores.
On average, CL AGN hosts have galaxy
S\'ersic indices and bulge-to-total-light ratios that are in the top $\approx$8\% of those of the reference AGN catalog.

\item  As galaxies quench, an evolutionary sequence has been identified for the transmutation of galaxies from SF through AGN to ultimately quiescence \citep{2009ApJ...705.1336C}. Our sample of CL AGNs are in the final stages of quenching and overlap most significantly with the AGN population of LINERs \citep{2019ApJ...883...31F}. This reinforces the idea that strong episodic bursts of accretion activity appear to occur preferentially as the amount of cold gas available for star formation and accretion dwindles. This is supported by the findings of \citet{2014MNRAS.444.3078G} that showed that lower luminosity AGNs display greater variability amplitudes \footnote{This is also supported by the anti correlation found between variability and Eddington ratio \citep{2004ApJ...609...69K,2010ApJ...716L..31A}.}. Within this evolutionary sequence, TDE hosts appear to represent the tail end of the galaxy quenching sequence. The enhanced nuclear stellar density in these systems appears to increase the probability of a stellar disruption \citepalias{2017ApJ...850...22L} and the lack of a strong accretion signature implies that TDE flares can be more readily detected.

\end{itemize}

The understanding of CL AGNs---the origin of the fueling gas, the mechanism by which it is transported to the SMBH and the nature of the accretion flow onto the BH---is a formidable challenge to theorists and to computational simulations of AGN accretion disks. It is, also, a daunting task for observers, in their search for detecting CL AGNs. To this end, a photometric criterion in galaxies has been identified in this study (as done for TDEs, \citetalias{2017ApJ...850...22L}), given by either the S\'ersic index or bulge fraction that could aid in anticipating CL AGN behavior.  
For forthcoming surveys detecting numerous nuclear transients per day, a photometric host galaxy selection criterion could be incredibly helpful for designing follow-up strategies aimed at characterizing CL AGNs.

\acknowledgments
We thank V. Baldassare, K. Bundy, J. Dai, D. French, M. Gaskell, B. Mockler, S., V. Pandya,  S. Raimundo and M. Vestergaard for useful discussions. We would also like to thank the referee for constructive comments that substantially helped improve the quality of the paper. We are indebted to S. Ellison for sharing with us so many key notions of quenching and galaxy transformation in the contemporary Universe. S.D, E.R-R and J.L-S are supported by the  Heising-Simons Foundation, the Danish National Research Foundation (DNRF132) and NSF (AST-1615881, AST-1911206 and AST-1852393). Parts of this research were supported by the Australian Research Council Centre of Excellence for All Sky Astrophysics in 3 Dimensions (ASTRO 3D), through project number CE170100013.

\appendix
\section{Disparity of Turn-on versus Turn-off CL AGN\lowercase{s}} \label{sec:appendix_a}
\begin{figure*}
\epsscale{0.6}
\plotone{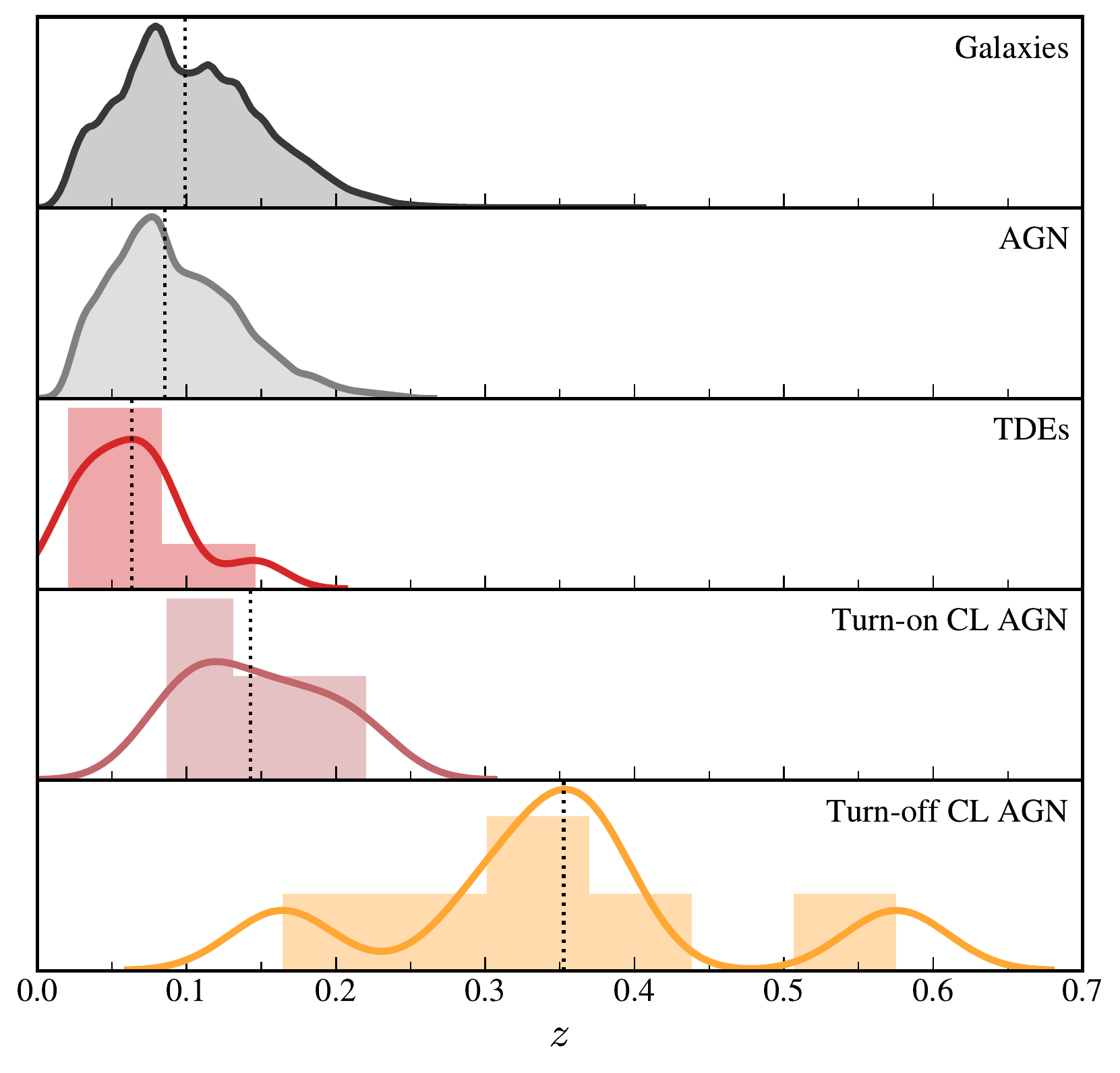}
\caption{Redshift distributions of galaxies from our reference catalog, AGN, TDEs, and samples of \textit{turn-on} and \textit{turn-off} CL AGN from \citet{2018ApJ...862..109Y}. There are 6 \textit{turn-off} and 15 \textit{turn-on} CL AGN. 11 of the \textit{turn-on} variety are used in our analysis of CL AGN hosts above.
\label{fig:figa1}}
\end{figure*}

As remarked in Section \ref{subsec:clagn_hosts}, we find that our CL AGN population is heavily biased toward {\it turn-on} types and we refer the reader to Table~\ref{tab:info} for our complete list of objects. In the paper from which we retrieve most of our sample, \citet{2018ApJ...862..109Y}, the authors report the discovery of a similarly unequal weighting of their {\it turn-on} versus {\it turn-off} counterparts: 15 of their 21 are {\it turn-on}. We consider two factors that could lead to this imbalance: biased discovery methods and a clear redshift  dependence. To the first point, detection methods frequently involve starting with a sample of galaxies and a sample of AGNs which then have the same variability constraints imposed upon them. Consequently, even if the incidence rate of CL behavior is intrinsically the same for AGNs and  galaxies, starting with order of magnitude differences in each group leads to unequal outcomes. As an example, for the photometric variability 
selection method used in \citet{2018ApJ...862..109Y} the authors start with 2,494,319 galaxies and 346,464 quasars before looking for significant  magnitude changes in both populations. The second point is clearly illustrated in Figure \ref{fig:figa1}: the median redshift of the {\it turn-off} sample is 0.3526 while the median redshift for the {\it turn-on} sample is 0.1427. Because our catalog is redshift limited (with a  median of $z = 0.099$), it is not surprising that we return a biased sample that leans heavily toward {\it turn-on} CL AGNs.

\section{The lower redshift ($\lowercase{z}< 0.15$) and higher redshift ($0.15<\lowercase{z}< 0.25$) CL AGN samples} \label{sec:appendix_b}
\begin{figure*}
\epsscale{0.6}
\plotone{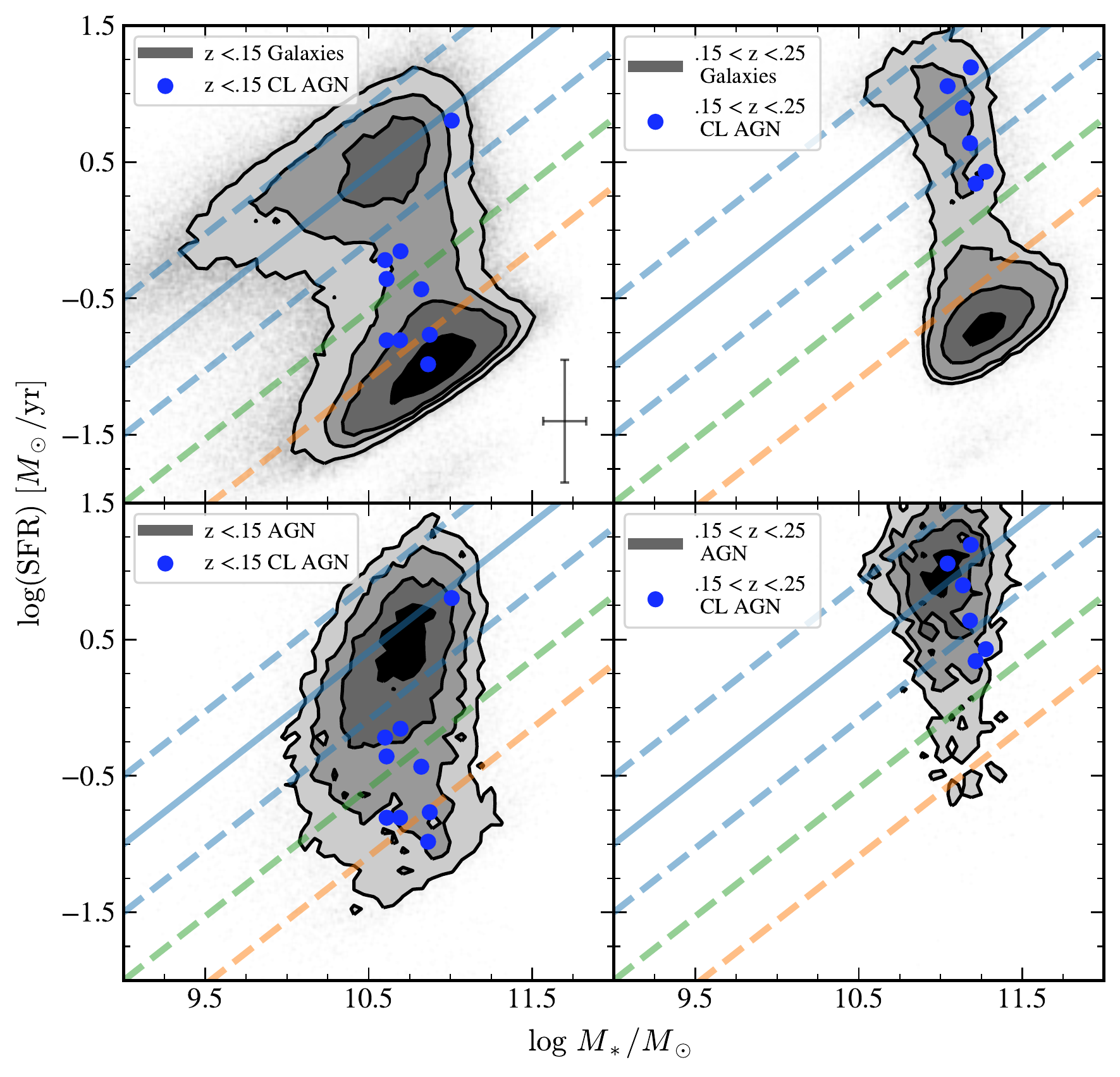}
\caption{SFR versus total stellar mass for galaxies (top row), AGN (bottom row), and CL AGN hosts (both). Similar to Figure~\ref{fig:fig1}.  Left side shows objects with redshifts $z<0.15$ while the right shows objects with redshifts $0.15<z<0.25$.
\label{fig:figa2}}
\end{figure*}

As a related metric of the nonconformity of CL AGN  host galaxies, we show total star formation rate versus total stellar mass for AGNs and galaxies in our reference catalog in Figure~\ref{fig:figa2}. The solid blue line describes the SFMS. As done in  Figure~\ref{fig:fig1} we assume a 1$\sigma$ scatter of 0.5 dex for the SFMS and delineate the green valley in this diagram as being 1$\sigma$ (blue dashed line) and 3$\sigma$ (orange dashed line) below the SFMS normalization. It is instantly evident from Figure~\ref{fig:figa2} that all but one of the CL AGN hosts in the lower ($z <0.15$) redshift range inhabit the green valley. On the contrary, the higher redshift CL AGN population ($0.15 < z < 0.25$) is more consistent with the AGN sample at a similar redshift range.  
However, it is clear that the location of the CL AGN hosts in these diagrams suggests that they could all be making a transition from the SFMS toward quiescence and that the lack of green valley CL AGN hosts located within the higher ($0.15 < z < 0.25$) redshift range could be explained by the scarcity of galaxies and AGN hosts in our catalog inhabiting this region (Figure~\ref{fig:figa2}). As such, we recognize that there could be a bias against CL AGN identification in green valley and dead elliptical galaxies in the higher redshift ($0.15 < z < 0.25$) range.

\section{CL AGN Host Galaxy Images} \label{sec:appendix_c}
\begin{figure*}
\epsscale{0.8}
\plotone{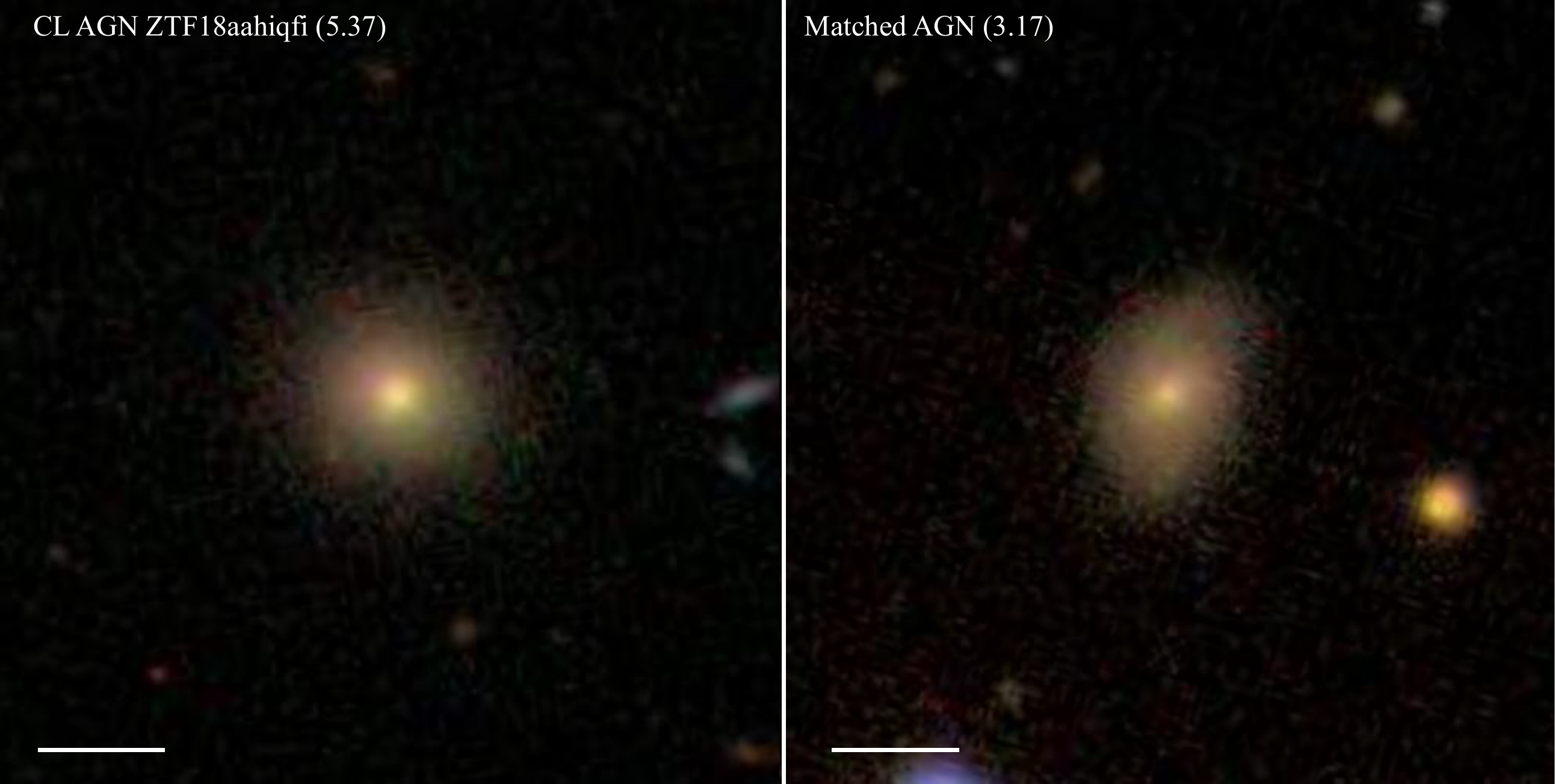}
\caption{SDSS images of CL AGN ZTF18aahiqfi (left) and a randomly selected AGN matched in SFR, $M_\ast$, and $z$ (right), illustrating the heightened central light concentration of the CL AGN host. The S\'ersic index of the AGN lies within the median value ($\pm 1 \sigma$) of the matched sample created for CL AGN ZTF18aahiqfi. The white bar in the images corresponds to $10^{''}$.
\label{fig:figa3}}
\end{figure*}

CL AGN host galaxies have more centrally concentrated light profiles as measured in S\'ersic index and bulge fraction than ``typical'' AGNs at their SMBH masses. As a
representative example of this, in Figure~\ref{fig:figa3}, we show the SDSS image of CL AGN ZTF18aahiqfi as well as a randomly selected AGN matched in SFR, stellar mass and redshift to the CL AGN host (7.5\% tolerance) but with a galaxy S\'ersic index within the median value ($\pm 1 \sigma$) of the matched sample. The high concentration of the CL AGN host is discernible by eye. The S\'ersic index is listed in parentheses in each galaxy image.

\section{Asymmetry Measurements and Galaxy Disturbances} \label{sec:appendix_d}

\begin{figure*}
\epsscale{1.1}
\plotone{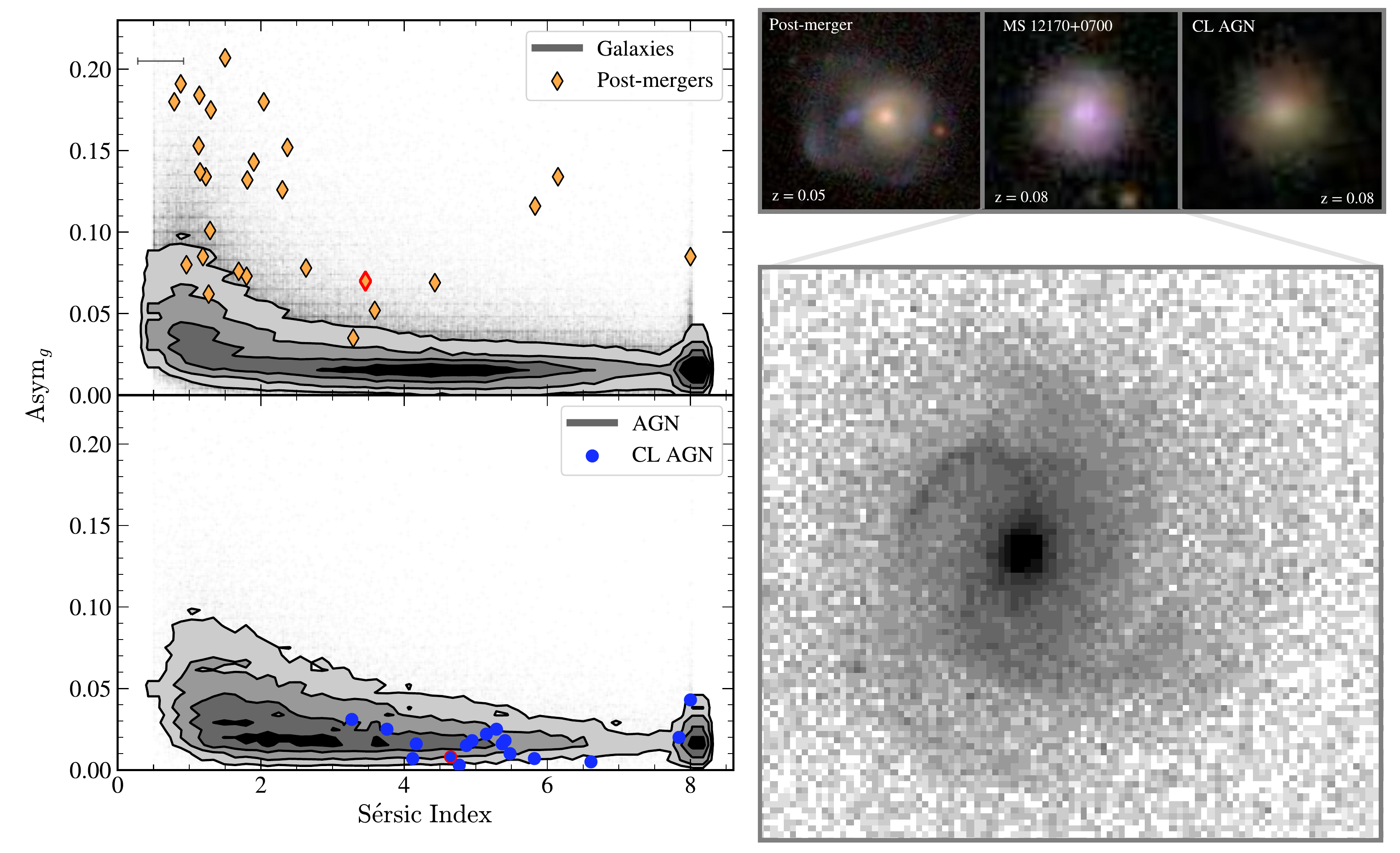}
\caption{\textit{Left column:} Asymmetry ($g$-band) versus S\'ersic index. The top panel shows our post-merger sample \citep{2013MNRAS.435.3627E} with galaxies as contours. The bottom panel shows CL AGN host galaxies with AGN contours. Median error in CL AGN S\'ersic index is shown in the top left (there is no significant error associated with asymmetry measurements in our catalog). The red-outlined galaxies in both plots are also featured in the upper right panel. \textit{Upper right panel:} SDSS images (left to right) of a post-merger galaxy, MS 12170+0700, and of CL AGN J1307+4506. Note that distorted features of the post-merger galaxy are clearly visible with SDSS resolution but that MS 12170+0700 and the CL AGN appear nearly featureless.  \textit{Lower right panel:} HST imaging of MS 12170+0700 from \citet{2003AJ....126.1690C}. The presence of a prominent non-axisymmetric
perturbation in the underlying mass distribution can be seen in the upper left.
\label{fig:figa4}}
\end{figure*}

The inward radial transport of gas in a galaxy requires torque on the gas that allows angular momentum to be transferred from the gas to the stars or dark
matter. Although this can be achieved in a swift and 
sensational way in a major merger or a strong tidal interaction, a slower but sizable mass inflow can still be driven by non-axisymmetric perturbations such as bars and spiral arms \citep{2014ARA&A..52..589H}.  For recent major mergers, the SDSS asymmetry indicators can be large, as shown in Figure~\ref{fig:figa4}. The asymmetry indicators are shown plotted against  S\'ersic index, which broadly increases for less asymmetric galaxies. The upper left plot shows galaxies as contours with the post-merger galaxies from \citet{2013MNRAS.435.3627E} depicted as diamonds, while the lower left plot has AGN as contours and CL AGN host galaxies as circles. 

Notably, our CL AGN host galaxies have small asymmetry indicators, suggesting that they are not the outcomes of recent major mergers.  The upper right panel of Figure~\ref{fig:figa4} shows the SDSS images of one post-merger galaxy (far left) and one CL AGN host (far right), both mid-range in terms of S\'ersic indices but with differing asymmetry measurements. These images illustrate that even a post-merger with a moderate asymmetry measurement exhibits  drastic features that can be clearly discerned with the SDSS resolution.  It is important to note that a small SDSS asymmetry indicator does not necessarily imply a lack of a sizable non-axisymmetric perturbation such as a bar or a spiral arm. The major limitation here is the SDSS resolution.  To illustrate this limitation, we show galaxy MS 12170+0700 (middle panel of top right panel and bottom right image of Figure~\ref{fig:figa4}). This galaxy, which has been matched in redshift to our CL AGN example, has a prominent spiral feature as seen with HST imaging \citep{2003AJ....126.1690C}. Contrasting this with the SDSS image for the same galaxy illustrates the limitations of our analysis when it comes to detecting fainter features. To this end, our analysis is consistent with the idea that CL AGN activation can be driven by non-axisymmetric perturbations in the underlying mass distribution that arise through internal dynamical processes in the cold disk.

\begin{deluxetable*}{ccccccccc}

\tablecaption{Table of CL AGNs\label{tab:info}}
\tablehead{
\colhead{Name} & \colhead{R.A.} & \colhead{Dec} & \colhead{Discovery Paper} & \colhead{Type} & \colhead{$z$} & \colhead{$\log\ M_{\rm bh}/M_\odot$} & \colhead{SFR} & \colhead{$\log\ M_\ast/M_\odot$} 
}
\startdata
J0831+3646 & 127.8844 & 36.7715 & \citet{2018ApJ...862..109Y} & Turn-on & 0.1951 & 8.589 & 0.6378 & 11.181 \\
J0909+4747 & 137.3835 & 47.7919 & \citet{2018ApJ...862..109Y} & Turn-on & 0.1171 & 7.348 & -0.2192 & 10.600 \\
ZTF18aaidlyq & 138.8794 & 48.2355 & \citet{2019ApJ...883...31F} & Turn-on & 0.1006 & 7.789 & -0.7665 & 10.874 \\
J0937+2602 & 144.3764 & 26.0423 & \citet{2018ApJ...862..109Y} & Turn-on & 0.1623 & 7.827 & 0.8964 & 11.137 \\
J1003+3525 & 150.8478 & 35.4177 & \citet{2018ApJ...862..109Y} & Turn-on & 0.1187 & 8.004 & -0.3574 & 10.609 \\
J1115+0544 & 168.9024 & 5.7471 & \citet{2018ApJ...862..109Y} & Turn-on & 0.0900 & 7.566 & -0.8071 & 10.610 \\
J1132+0357 & 173.1214 & 3.9581 & \citet{2018ApJ...862..109Y} & Turn-on & 0.0910 & 7.932 & -0.8068 & 10.692 \\
ZTF18aasuray & 173.4831 & 67.0186 & \citet{2019ApJ...883...31F} & Turn-on & 0.0397 & 8.167 & -0.4329 & 10.822 \\
ZTF18aasszwr & 186.4596 & 51.1462 & \citet{2019ApJ...883...31F} & Turn-on & 0.1680 & 8.617 & 0.3415 & 11.215 \\
ZTF18aahiqfi & 193.5158 & 49.2480 & \citet{2019ApJ...883...31F} & Turn-on & 0.0670 & 8.257 & -0.9826 & 10.864 \\
J1319+6753 & 199.8781 & 67.8987 & \citet{2018ApJ...862..109Y} & Turn-on & 0.1664 & 7.507 & 1.056 & 11.043 \\
J1447+2833 & 221.9760 & 28.5567 & \citet{2018ApJ...862..109Y} & Turn-on & 0.1631 & 7.999 & 1.1936 & 11.186 \\
J1545+2511 & 236.3735 & 25.1911 & \citet{2018ApJ...862..109Y} & Turn-on & 0.1171 & 8.124 & 0.8033 & 11.008 \\
J1550+4139 & 237.5718 & 41.6506 & \citet{2018ApJ...862..109Y} & Turn-on & 0.2203 & 8.265 & 0.4288 & 11.277 \\
J1552+2737 & 238.2429 & 27.6246 & \citet{2018ApJ...862..109Y} & Turn-on & 0.0866 & 8.086 & -0.1545 & 10.695 \\
J1307+4506 & 196.8208 & 45.1126 & \citet{2020ApJ...889...46S} & Turn-off & 0.0840 & N/A & N/A & 10.261 \\
J1428+1723 & 217.1946 & 17.3981 & \citet{2020ApJ...889...46S} & Turn-off & 0.1037 & N/A & N/A & 10.626 \\
\enddata
\end{deluxetable*}

\begin{deluxetable*}{ccccc}

\tablecaption{Structural Properties of CL AGNs \label{tab:struc}}
\tablehead{
\colhead{Name} & \colhead{S\'ersic Index} & \colhead{B/T\tablenotemark{a}} & \colhead{Scale\tablenotemark{b}} & \colhead{$R_{\rm hl}$\tablenotemark{c}}  
}
\startdata
J0831+3646 & 5.15 $\pm$ 0.46 & 0.29 $\pm$ 0.03 & 3.236 & 6.24  \\
J0909+4747 & 7.84 $\pm$ 0.33 & 0.67 $\pm$ 0.04 & 2.118 & 3.04 \\
ZTF18aaidlyq & 6.61 $\pm$ 0.12 & 0.68 $\pm$ 0.01 & 1.853 & 6.44 \\
J0937+2602 & 5.41 $\pm$ 0.63 & 1.00 $\pm$ 0.01 & 2.790 & 6.28 \\
J1003+3525 & 4.77 $\pm$ 0.40 & 0.85 $\pm$ 0.07 & 2.143 & 1.53 \\
J1115+0544 & 4.12 $\pm$ 0.22 & 0.43 $\pm$ 0.04 & 1.679 & 3.14 \\
J1132+0357 & 4.87 $\pm$ 0.32 & 0.96 $\pm$ 0.03 & 1.696 & 2.19 \\
ZTF18aasuray & 3.76 $\pm$ 0.02 & 0.42 $\pm$ 0.02 & 0.786 & 7.64 \\
ZTF18aasszwr & 5.48 $\pm$ 0.39 & 1.00 $\pm$ 0.01 & 2.870 & 5.34 \\
ZTF18aahiqfi & 5.37 $\pm$ 0.10 & 0.71 $\pm$ 0.04 & 1.283 & 5.18 \\
J1319+6753 & 4.17 $\pm$ 0.19 & 0.70 $\pm$ 0.05 & 2.848 & 3.87 \\
J1447+2833 & 8.00 $\pm$ 0.03 & 0.44 $\pm$ 0.05 & 2.802 & 10.44 \\
J1545+2511 & 4.95 $\pm$ 0.17 & 0.69 $\pm$ 0.02 & 2.118 & 3.59 \\
J1550+4139 & 5.29 $\pm$ 0.43 & 0.51 $\pm$ 0.04 & 3.556 & 7.89 \\
J1552+2737 & 3.27 $\pm$ 0.11 & 0.17 $\pm$ 0.01 & 1.621 & 6.84 \\
J1307+4506 & 4.65 $\pm$ 0.58 & 0.37 $\pm$ 0.04 & 1.578 & 2.82 \\
J1428+1723 & 5.82 $\pm$ 0.35 & 0.90 $\pm$ 0.05 & 1.905 & 1.39 \\
\enddata
\tablenotetext{a}{B/T: Bulge-to-total-light ratio in $g$ band.}
\tablenotetext{b}{Physical scale in arcsec/kpc at redshift z.}
\tablenotetext{c}{Galaxy circular half-light radius in kpc and in $g$ band.}
\end{deluxetable*}

\end{document}